\documentclass{article}
\usepackage{spconf,amsmath,graphicx,hyperref}
\usepackage[table]{xcolor}
\usepackage{booktabs}
\usepackage{pifont}
\usepackage{makecell}
\usepackage{multirow}
\newcommand{\cmark}{\ding{51}}%
\newcommand{\xmark}{\ding{55}}%

\title{LJ-Spoof: A Generatively Varied Corpus for Audio Anti-Spoofing and Synthesis Source Tracing}
\name{ Surya Subramani, Hashim Ali, Hafiz Malik}
\address{Electrical and Computer Engineering\\
        University of Michigan\\
	4901 Evergreen Rd, Dearborn, MI 48128, USA}
\begin{document}
%
\maketitle
\begin{abstract}
Speaker-specific anti-spoofing and synthesis-source tracing are central challenges in audio anti-spoofing. Progress has been hampered by the lack of datasets that systematically vary model architectures, synthesis pipelines, and generative parameters. To address this gap, we introduce \textbf{LJ-Spoof}, a speaker-specific, generatively diverse corpus that systematically varies prosody, vocoders, generative hyperparameters, \textit{bona fide} prompt sources, training regimes, and neural post-processing. The corpus spans one speakers-including studio-quality recordings-30 TTS families, 500  generatively variant subsets, 10 \textit{bona fide} neural-processing variants, and more than 3 million utterances. This variation-dense design enables robust speaker-conditioned anti-spoofing and fine-grained synthesis-source tracing. We further position this dataset as both a practical reference training resource and a benchmark evaluation suite for anti-spoofing and source tracing.
\end{abstract}

\section{Introduction}
\label{sec:intro}
LLM-era speech synthesis has reached a level of naturalness that often fools human listeners. Advances in self-supervised learning and discrete speech tokenization, together with powerful acoustic encoders that capture nonsemantic cues (speaker identity, style, prosody, emotion), have further narrowed the gap between synthetic and bona fide speech. While these gains benefit voice applications, they also lower the barrier to misuse\cite{knibbs_researchers_2024}: open-source systems can clone a voice from a 5--10\,s reference and generate convincing speech within minutes, enabling fraud and opinion manipulation at scale.

Beyond broad, cross-speaker detection, protecting \emph{individual} voices is crucial. For everyday users, cloned voices can be weaponized against family or coworkers; for public figures, they can shape narratives and sway public opinion. This motivates \emph{speaker-specific} anti-spoofing as a complementary and urgently needed research direction.

Surveys (e.g., \cite{tan2021surveyneuralspeechsynthesis}) summarize the neural TTS landscape, and recent work on controllable TTS \cite{2024arXiv241206602X} shows that modern generators can produce numerous variants from the same prompt by adjusting parameters (e.g., temperature, solver steps, duration/speed) and configurations. A central question follows: \emph{Are current anti-spoofing systems robust to these generative variations? Can they distinguish semantically/prosodically manipulated fakes from neurally post-processed or resynthesized bona fides, and from untouched bona fides?} Addressing these questions requires a careful look at existing anti-spoofing datasets the starting point for model development and evaluation and motivates the analysis and dataset design presented in this work.

\section{Audio Anti-Spoofing Datasets and Shortcomings}
Existing corpora span single-speaker TTS \cite{firc2024diffuse,frank2021wavefake}, multi-speaker TTS/VC \cite{wang2024asvspoof5,yamagishi2021asvspoof2021,yamagishi2019asvspoof2019db,ali25_interspeech}, multilingual \cite{muller2024mlaad}, diffusion-focused \cite{bhagtani2024diffssd,du2024dfadd}, and codec-based \cite{du2025codecfake}. The ASVspoof series steadily raises attack realism but still lacks sufficient bona fide \emph{and} spoofed material per speaker for rigorous single-speaker analysis. \textbf{CodecFake-Omni} \cite{du2025codecfake} is the largest neural-codec set (trained on 31 codec families; evaluates 17 CoSG models). Diffusion-centric corpora are complementary: \textbf{DFADD} \cite{du2024dfadd} trains on five diffusion/flow-matching generators (EN/ZH) and boosts EERs over ASVspoof-only baselines; \textbf{DiffSSD} \cite{bhagtani2024diffssd} aggregates ten diffusion systems, reveals detector blind spots, and shows retraining recovers performance. \textbf{Diffuse or Confuse} \cite{firc2024diffuse} (LJSpeech) offers four diffusion syntheses plus diffusion re-vocoding of non-diffusion fakes, emphasizing curation and analysis over full countermeasure training.

Single-speaker corpora include \textbf{WaveFake} \cite{frank2021wavefake} (re-vocoded bona fide via neural vocoders) and \textbf{Diffuse or Confuse} (diffusion acoustic + re-vocoding). \textbf{MLAADv5} \cite{muller2024mlaad} compiles 91 subsets from 42 architectures across 38 languages, later grouped into 27 TTS families; spoofs are not speaker-specific, prioritizing multilingual coverage. \textbf{STOPA} \cite{firc2025stopa} targets open-world source tracing with 700k samples from 13 synthesizers (VCTK-based).

\begin{table}[t]
\centering
\begingroup
\tiny
\setlength{\tabcolsep}{4pt}
\renewcommand{\arraystretch}{1.0}
\providecommand{\cmark}{\checkmark}
\providecommand{\xmark}{--}
\begin{tabular}{@{}l l r c c c c c r r@{}}
\toprule
Dataset & Type & TTS & VocVar & GenVar & ReVoc & ReCod & FWVar & Subsets & Synth \\
\midrule
TIMIT\mbox{-}TTS        & Multi   & 12 & \xmark & \xmark & \xmark & \xmark            & \xmark &  12 &    79k \\
FoR                      & Multi   &  7 & \xmark & \xmark & \xmark & \xmark            & \xmark &   7 &    87k \\
In-the-Wild              & Multi   & -- & --     & --     & --     & --                & --     &  -- &    11k \\
ASVspoof2019             & Multi   &  6 & \cmark & \xmark & \xmark & \xmark            & \xmark &  17 &   108k \\
ASVspoof2021             & Multi   &  6 & \cmark & \xmark & \xmark & \cmark            & \xmark &  17 &   148k \\
ASVspoof 5               & Multi   & 15 & \cmark & \xmark & \xmark & \cmark            & \xmark &  32 & 1{,}211k \\
CodecFake\mbox{-}Omni    & Multi   & 17 & \xmark & \xmark & \xmark & \cmark$^{\dagger}$& \xmark &  17 &    --  \\
DFADD                    & Multi   &  5 & \xmark & \xmark & \xmark & \xmark            & \xmark &   5 &   163k \\
DiffSSD                  & Multi   & 10 & \xmark & \xmark & \xmark & \xmark            & \xmark &  10 &    70k \\
WaveFake                 & Single  &  0 & \xmark & \xmark & \xmark & \xmark            & \xmark &  10 &   118k \\
Diffuse/Confuse          & Single  &  8 & \xmark & \xmark & \cmark & \xmark            & \xmark &  12 &   183k \\
STOPA                    & Multi   &  6 & \cmark & \xmark & \xmark & \xmark            & \cmark &  13 &   699k \\
MLAAD                    & Multi   & 27 & \xmark & \xmark & \xmark & \xmark            & \cmark &  91 &   175k \\
FamousFigures                    & Multi   & 10 & \xmark & \xmark & \xmark & \xmark            & \cmark &  10 &   265k \\
\midrule
\textbf{LJ\mbox{-}Spoof} & \textbf{Single} & \textbf{30} & \textbf{\cmark} & \textbf{\cmark} & \textbf{\cmark} & \textbf{\cmark} & \textbf{\cmark} & \textbf{500} & \textbf{2{,}800k} \\
\bottomrule
\end{tabular}
\endgroup
\caption{Anti-spoofing datasets and synthesis characteristics. Abbrev.: VocVar = vocoder-based; GenVar = generative-parameter; ReVoc = re-vocoding; ReCod = codec resynthesis; FWVar = framework-based. $^{\dagger}$Bona fide only.}
\label{tab:datasets}
\vspace{-1mm}
\end{table}

Overall, these datasets advance specific axesmulti-speaker, single-speaker, diffusion, codec, multilingual yet none provide a controlled, comprehensive \emph{single-speaker} resource that systematically varies architectures, generative settings, re-vocoding, and codec resynthesis while holding external factors fixed. The table \ref{tab:datasets} clearly portrays the difference between the existing datasets and our novel LJ-Spoof

Our study introduces the novel dataset \textbf{LJ-Spoof}, which features:
\begin{itemize}
  \item \textbf{33 TTS families:} 11 single-speaker (SS-TTS), 2 fine-tunable multi-speaker (FT-TTS), and 20 zero-shot (ZS-TTS), covering state-of-the-art systems and making the corpus uniquely focused on TTS-synthesized spoofs. Families span all major paradigms from basic RNN to LLM-based architectures.
  \item \textbf{Training regimes captured:} 11 SS-TTS families pretrained on the target speaker's bona fide audio, 2 FT-TTS families, and 20 ZS-TTS families jointly isolating the effects of single-speaker, fine-tuning, and zero-shot synthesis.
  \item \textbf{~500 variant subsets:} Systematically varied by training regime, synthesis input, generative parameters, acoustic--vocoder pairing, and neural post-processing, yielding a breadth and density of configurations rarely explored in prior work.
  \item \textbf{Reproducible protocol:} A detailed, versioned protocol file documents architectural choices, configuration settings, and neural post-processing steps, enabling rigorous, fine-grained single-speaker anti-spoofing analyses by the community.
\end{itemize}

Table \ref{tab:datasets} clearly portrays that LJ-Spoof is a only Single Speaker dataset which is large and synthetically diverse with Vocoder-based-variations, Generative Parameter Variations, Re-Vocoding, Re-Codec(Codec Resynthesis) and Framework based Variations.

\section{LJ-Spoof}
\begin{figure*}[ht]
    \centering
    \includegraphics[width=\linewidth]{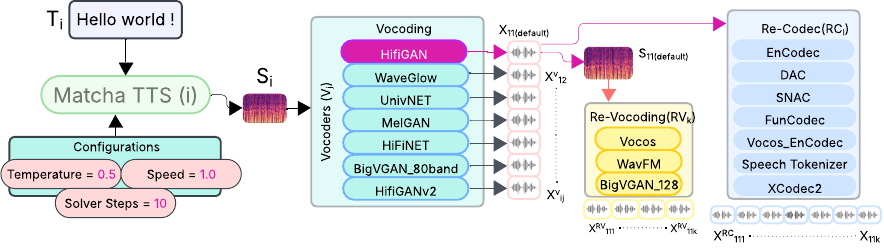}
    \caption{Dataset Generation Workflow for Matcha-TTS. A mel-spectrogram $S$ is synthesized from text $T$ using default generative parameters (sampling temperature $\tau=0.5$, peak rate $R=1.0$, ODE solver steps $D=10$). $S$ is then decoded by multiple vocoders (blue), including the default HiFiGAN (purple), to produce diverse waveform subsets. Finally, the default waveform undergoes post-processing via re-vocoding (yellow) and re-codec (blue).}
    \label{fig:matcha}
\end{figure*}

In this section we summarize the construction of the \emph{LJ-Spoof} dataset, the design choices behind it, and the analysis objectives.

\subsection{Bonafide and Default Subsets}
We aim to isolate synthesizer and parameter effects under a single-speaker, studio-quality setting. LJSpeech \cite{ljspeech17} contains 13{,}100 utterances (avg.~6.57\\,s) from a single narrator (\mbox{$\approx$24\,h} total), supporting SS-TTS training, FT-TTS fine-tuning, and ZS-TTS reference prompts. We treat these 13k utterances as the default \textit{bona fide} set. Using this default set, we generate 34 default synthetic subsets across 30 TTS families, each under its published or suggested configurations. 
\subsection{Variant Subsets}
 Beyond this 35 default subsets, using training categories, acoustic-vocoder pairing, reference utterance and Text inputs, generative parameters and Neural post processing variations we will be creating more than 500 synthetic subsets like mentioned in Table 2. We used only the official repositories, Coqui TTS and Nvidia's Nemo for implementing both default and variant subsets.
 \subsubsection{Training Categories Variants}
 Training categories refer to the Text-to-Speech training method. It can specifically pretrained for one speaker (SS-TTS)\cite{mehta2024matcha}, pretrained on Multispeaker with the capability of finetuning its speaker encoder for one speaker(FT-TTS)\cite{casanova2024xtts} and Zeroshot inference capable models pretrained on large multispeaker datasets (ZS-TTS)\cite{chen2024f5tts}. Mostly each TTS family are trained for specific training categories except \cite{li2023styletts2} which has all three training categories weights and \cite{casanova2024xtts} has FT-TTS and ZS-TTS.

\subsubsection{Acoustic--Vocoder Variants}
Most SS-TTS families use acoustic models pretrained on LJSpeech, which is a de facto ``go-to'' corpus for SS-TTS implementations. These acoustic models are typically cascaded with neural vocoders \cite{yang2023hifi} that also provide open-source LJSpeech-pretrained weights. Such a cascaded design with independently pretrained components allows flexible mix-and-match of acoustic and vocoder modules, provided the vocoder supports the mel-spectrogram configuration (e.g., number of mel bands, frequency range) emitted by the acoustic model. A small number of pairings are unsupported due to mismatches between the acoustic model's mel output specification and the vocoder's expected mel input configuration.
\subsubsection{Input Variants}
Most FT-TTS (fine-tuned) and ZS-TTS (zero-shot) models take two inputs: (i) a \emph{prompt audio} that provides speaker or style or acoustic reference, and (ii) a \emph{target text} to be synthesized. We systematically vary both inputs to construct the following subsets.

\textbf{Prompt-audio} variants comprise three subsets: (i) \emph{Fixed Prompt}, where a single prompt audio is reused for the entire subset; (ii) \emph{Random Prompt}, where a prompt audio with content different from the target text is selected independently for each synthesis; and (iii) \emph{Same-Content Prompt}, where the prompt audio matches the target text. Unless otherwise noted, the default synthetic subsets use the Fixed Prompt.

\textbf{Target-text} variants comprise two subsets: (i) transcriptions of 13k bona fide LJSpeech utterances, and (ii) 5k LLM-generated transcripts covering diverse topics (Technology, Entertainment, Geography, History, and Politics). Both the default and variant synthetic subsets contain 18k utterances drawn from the union of these bona fide and generated texts.

\subsubsection{Generative-Parameter Variants}
We introduce generative-parameter variants for each TTS family by systematically sweeping family-specific inference controls. Concretely, we vary: (i) \emph{stochastic initialization} (e.g., random-noise seeds for diffusion-based models); (ii) \emph{inference steps}, including diffusion or ODE solver step counts that govern the noise-removal trajectory; (iii) \emph{sampling temperature} $\tau \in [0.1, 1.0]$, which modulates output randomness; (iv) \emph{expressiveness/emotion scaling} $\alpha \in [0,1]$ or discrete emotion labels, controlling style intensity; and (v) \emph{duration/speed factor} $r$ (with $r<1$ slower and $r>1$ faster than the default). These controlled variations yield families of synthetic subsets that isolate the effect of each parameter on the resulting speech.

While the above are general controls, individual TTS families expose model-specific knobs. For example, \cite{li2023styletts2} does not use explicit emotion labels or a conventional temperature; instead, it employs $(\alpha,\beta)$ for controlling how much emotion/style is captured from the prompt audio and an \emph{embedding-scale} parameter that effectively governs sampling variability.
\subsubsection{Neural Post-Processing Variants}
We also apply neural post-processing to the bona fide audio and the 34 default synthetic subsets using five state-of-the-art vocoder and codec models. Two variants are considered: (i) \emph{re-vocoding}, where each utterance is converted to a mel-spectrogram and resynthesized with a neural vocoder; and (ii) \emph{codec resynthesis} (re-CODEC), where discrete codes extracted from the utterance are decoded by a neural codec to reconstruct the waveform. This variants produces 10 bonafide subsets and 340 synthetic subsets. Figure \ref{fig:matcha} explains the clear pipeline of how different generative acoustic model configuration and vocoders paired to produce different acoustic-vocoder variants and neurally post processing the default configuration and architecture.

\subsection{Metadata Preparation}
Given the dataset's scale and complexity, metadata preparation is critical. We provide two complementary metadata files that share a common key column called ``UniqueID.'' The first file (file index) lists, for each sample, the UniqueID, the absolute or relative file path, and the top-level label (bonafide or spoof). The second file (variant index) links each UniqueID to its generative details, including the variant category (e.g., Training-based, Architecture-based, input-based), the variant label (a unique identifier for the specific variant), a short variant description(e.g., Temperature = 0.5), the TTS model name, the TTS model type (e.g., Diffusion, LLM, Transformer, DiT), and a flag indicating whether the sample belongs to a default subset or a variant subset.

\section{Analysis Scope and Discussion}
Following data generation, we delineate the scope of analyses enabled by this large single-speaker corpus and highlight its advantages for controlled, fine-grained study. Our analysis centers on three orthogonal axes of variation \emph{training category} (SS-/FT-/ZS-TTS), \emph{input source} (e.g., bona fide scripts vs.\ generated texts, fixed vs.\ random references), and \emph{generative parameters} (e.g., sampling temperature, diffusion/solver steps, duration/speed controls) as well as \emph{neural post-processing} (re-vocoding and codec resynthesis).
\begin{table}[t]
\centering
\setlength{\tabcolsep}{4pt}
\renewcommand{\arraystretch}{0.95}
\label{tab: tts-variations-subsets}
\resizebox{\linewidth}{!}{
\begin{tabular}{lccc c cc ccccc cc c}
\toprule
\multirow{2}{*}{TTS Family} & \multicolumn{3}{c}{Training} & \multirow{2}{*}{DS} & \multicolumn{2}{c}{Input} & \multicolumn{5}{c}{Architecture} & \multicolumn{2}{c}{Neural post} & \multirow{2}{*}{Total} \\
 & SS & FT & ZS &  & prom. & Text & Voc. & Steps & Temp & Emo & Dur & Re\mbox{-}Voc. & Re\mbox{-}COD. &  \\
\midrule
Neural HMM     & \cmark & \xmark & \xmark & 1 & \xmark & \xmark & 4 & 5 & \xmark & \xmark & \xmark & 5 & 5 & 19 \\
Tacotron2      & \cmark & \xmark & \xmark & 1 & \xmark & \xmark & 4 & 7 & \xmark & \xmark & \xmark & 5 & 5 & 21 \\
FastSpeech2    & \cmark & \xmark & \xmark & 1 & \xmark & \xmark & 4 & 5 & \xmark & \xmark & \xmark & 5 & 5 & 19 \\
SpeedySpeech   & \cmark & \xmark & \xmark & 1 & \xmark & \xmark & 4 & 5 & \xmark & \xmark & \xmark & 5 & 5 & 19 \\
FastPitch      & \cmark & \xmark & \xmark & 1 & \xmark & \xmark & 4 & 7 & \xmark & \xmark & \xmark & 5 & 5 & 21 \\
Overflow       & \cmark & \xmark & \xmark & 1 & \xmark & \xmark & 4 & 5 & \xmark & \xmark & \xmark & 5 & 5 & 19 \\
Mixer\mbox{-}TTS    & \cmark & \xmark & \xmark & 1 & \xmark & \xmark & 4 & 4 & \xmark & \xmark & \xmark & 5 & 5 & 18 \\
Glow\mbox{-}TTS     & \cmark & \xmark & \xmark & 1 & \xmark & \xmark & 4 & 5 & \xmark & \xmark & \xmark & 5 & 5 & 19 \\
VITS            & \cmark & \xmark & \xmark & 1 & \xmark & \xmark & 4 & 1 & \xmark & \xmark & \xmark & 5 & 5 & 15 \\
STYLETTS2       & \cmark & \cmark & \cmark & 3 & 3 & 4 & 2 & 2 & 2 & 3 & 2 & 5 & 5 & 78 \\
Matcha\mbox{-}TTS   & \cmark & \xmark & \xmark & 1 & \xmark & \xmark & 4 & 4 & 2 & 3 & \xmark & 5 & 5 & 23 \\
YourTTS         & \xmark & \xmark & \cmark & 1 & 3 & 4 & \xmark & \xmark & 3 & \xmark & \xmark & 5 & 5 & 20 \\
Tortoise        & \xmark & \xmark & \cmark & 1 & 3 & 4 & \xmark & 4 & \xmark & \xmark & \xmark & 5 & 5 & 21 \\
VALL\mbox{-}E        & \xmark & \xmark & \cmark & 1 & 3 & 4 & \xmark & \xmark & 3 & \xmark & \xmark & 5 & 5 & 20 \\
WhisperSpeech   & \xmark & \xmark & \cmark & 1 & 3 & 4 & \xmark & \xmark & 3 & \xmark & \xmark & 5 & 5 & 20 \\
XTTSv2          & \xmark & \cmark & \cmark & 2 & 3 & 4 & \xmark & \xmark & 3 & \xmark & 3 & 5 & 5 & 46 \\
HierSpeech++    & \xmark & \xmark & \cmark & 1 & 3 & 4 & \xmark & \xmark & 3 & \xmark & \xmark & 5 & 5 & 20 \\
MaskGCT         & \xmark & \xmark & \cmark & 1 & 3 & 4 & \xmark & \xmark & 3 & \xmark & 3 & 5 & 5 & 23 \\
FireRedTTS      & \xmark & \xmark & \cmark & 1 & 3 & 4 & \xmark & \xmark & 3 & \xmark & 3 & 5 & 5 & 23 \\
F5TTS           & \xmark & \xmark & \cmark & 1 & 3 & 4 & \xmark & 3 & 3 & 3 & \xmark & 5 & 5 & 29 \\
E2TTS           & \xmark & \xmark & \cmark & 1 & 3 & 4 & \xmark & 3 & 3 & 3 & \xmark & 5 & 5 & 29 \\
FishSpeech      & \xmark & \xmark & \cmark & 1 & 3 & 4 & \xmark & \xmark & 3 & 3 & \xmark & 5 & 5 & 23 \\
CosyVoice2      & \xmark & \xmark & \cmark & 1 & 3 & 4 & \xmark & \xmark & 3 & 3 & \xmark & 5 & 5 & 23 \\
Zonos           & \xmark & \xmark & \cmark & 1 & 3 & 4 & \xmark & \xmark & 3 & 3 & 3 & 5 & 5 & 26 \\
Spark\mbox{-}TTS     & \xmark & \xmark & \cmark & 1 & 3 & 4 & \xmark & \xmark & 3 & 3 & 3 & 5 & 5 & 26 \\
MegaTTS3        & \xmark & \xmark & \cmark & 1 & 3 & 4 & \xmark & 3 & 3 & \xmark & \xmark & 5 & 5 & 23 \\
OrpheusTTS      & \xmark & \xmark & \cmark & 1 & 3 & 4 & \xmark & \xmark & 3 & \xmark & 3 & 5 & 5 & 23 \\
MuyanTTS        & \xmark & \xmark & \cmark & 1 & 3 & 4 & \xmark & \xmark & 3 & \xmark & 3 & 5 & 5 & 23 \\
OuteTTS         & \xmark & \xmark & \cmark & 1 & 3 & 4 & \xmark & \xmark & 3 & \xmark & 3 & 5 & 5 & 23 \\
ChatterboxTTS   & \xmark & \xmark & \cmark & 1 & 3 & 4 & \xmark & \xmark & 3 & 3 & 3 & 5 & 5 & 26 \\
IndexTTS2       & \xmark & \xmark & \cmark & 1 & 3 & 4 & \xmark & \xmark & 3 & 3 & 3 & 5 & 5 & 26 \\
\bottomrule
\end{tabular}}
\caption{TTS families and per-family variant subset counts. Columns summarize training regime (SS/FT/ZS), default-subset count(DS), input conditions, architectural/generative controls, neural post-processing, and the resulting total number of variant subsets per family. \emph{Legend:} prom.~= prompt voice; Text = target text; Voc.~= acoustic--vocoder variants; Steps = diffusion steps; Temp = sampling temperature; Emo = emotion; Dur = duration; Re\mbox{-}Voc.~= re-vocoded; Re\mbox{-}COD.~= re-CODEC.}
\end{table}
\subsection{Why Variants?}
Input-based variants particularly pairing generated texts with bona fide scripts encourage detectors to generalize across both authentic and manipulated linguistic content, rather than overfitting to samples containing only the ``originally spoken words.'' 

Training-category variants help disentangle artifacts among models pretrained on a single speaker (SS-TTS), models fine-tuned for the target speaker (FT-TTS), and zero-shot systems (ZS-TTS) that condition on a previously unseen speaker using few seconds reference clip. 

Generative-parameter variants expose detectors to multiple realizations of the same synthesizer similar in architecture but different in configuration so they do not overfit to a default setting (e.g., temperature = 0.7). This is essential because adversaries can adjust such parameters to produce more natural-sounding spoofs which can influence people's opinion. 

While \cite{du2025codecfake,xie2024codecfake} argue for treating CODEC--resynthesized bona fide speech i.e., neurally post-processed but otherwise unaltered audio as spoof in order to capture CODEC artifacts, we take a different stance: we label CODEC-based TTS, and resynthesized or re-vocoded \emph{synthetic} speech as spoof, and we keep CODEC--resynthesized and re-vocoded \emph{bona fide} speech as bona fide. The core anti-spoofing problem concerns speech that is contextually, acoustically, emotionally, or vocally manipulated; none of these attributes are changed by mere resynthesis of a bona fide signal. This labeling choice reduces bias toward artifacts introduced by mel-spectrogram or acoustic-code decompression, and instead encourages detectors to focus on the distortions and anomalies that arise when constructing these representations for \emph{manipulated} content.
\subsection{Split Protocol}
To train contemporary anti-spoofing systems, we sample equal numbers of bona fide and spoof utterances. We propose a strategic split that allocates 750 utterances per ZS-TTS and SS-TTS while ensuring coverage across variant families. Rows correspond to text-similarity conditions: \textbf{TTT} (identical text in bona fide and generated spoofs), \textbf{TT} (identical text across TTS default subsets), \textbf{T} (identical text across variant subsets), and \textbf{t} (all texts different across labels, TTS families, and variants). For bonafides, the split balances original and neurally post-processed speech, with some items sharing text across labels and others using content not reused to create spoofs. 

\section{Conclusion and Future Scope}
This work introduces a ``go-to'' evaluation corpus for assessing the robustness of generalized anti-spoofing systems against diverse synthesis attacks, provides training data for speaker-specific anti-spoofing architectures, and offers an ideal resource for synthesis-source tracing enabled by the breadth of spoof variants and ample utterance counts per subset. The dataset will be open-sourced to support reproducible research and broad community adoption.

We will extend the corpus and analyses to in-the-wild speech and to multiple speakers, enabling deeper, more comprehensive investigations into the effects of individual synthesizers and recording conditions.

\bibliographystyle{IEEEbib}
\bibliography{strings,refs}

\end{document}